\documentclass[a4paper, amsfonts, amssymb, amsmath, reprint, showkeys, nofootinbib, superscriptaddress, twoside, longbibliography]{revtex4-1}
\usepackage[english]{babel}
\usepackage[utf8]{inputenc}
\usepackage{amsthm}
\usepackage{mathtools}
\usepackage{physics}
\usepackage{xcolor}
\usepackage{graphicx}
\usepackage[left=23mm,right=13mm,top=35mm,columnsep=15pt]{geometry} 
\usepackage{adjustbox}
\usepackage{placeins}
\usepackage[T1]{fontenc}
\usepackage{lipsum}
\usepackage{csquotes}
\usepackage{hyperref}
\usepackage{soul}
\usepackage{changes}
\usetikzlibrary{shapes.misc,arrows}
\usepackage{siunitx}

\begin{document}
    \title{A Snapshot of Relativistic Motion: Visualizing the Terrell Effect}


\author{Dominik Hornof $^*$}
\affiliation{Vienna Center for Quantum Science and Technology, Atominstitut, TU Wien, Vienna, Austria}
\affiliation{University Service Centre for Transmission Electron Microscopy,TU Wien, Wiedner Hauptstraße 8-10/E057-02, 1040 Wien, Austria}
\author{Victoria Helm $^*$}
 \affiliation{University of Vienna, Faculty of Physics, VCQ, 
A-1090 Vienna, Austria}
  \affiliation{University of Vienna, Max Perutz Laboratories, 
Department of Structural and Computational Biology, A-1030 Vienna, Austria}
\affiliation{Vienna Center for Quantum Science and Technology, Atominstitut, TU Wien, Vienna, Austria}
\author{Enar de Dios Rodriguez}
\affiliation{Kunstuniversität Linz, 
Hauptplatz 8,
4010 Linz, Austria}
\affiliation{IFK, Internationales Forschungszentrum
Kulturwissenschaften
Kunstuniversität Linz in Wien, Reichsratsstraße 17, 1010 Wien, Austria}
  \author{Thomas Juffmann}
  \affiliation{University of Vienna, Faculty of Physics, VCQ, 
A-1090 Vienna, Austria}
  \affiliation{University of Vienna, Max Perutz Laboratories, 
Department of Structural and Computational Biology, A-1030 Vienna, Austria}
\author{Philipp Haslinger}
\affiliation{Vienna Center for Quantum Science and Technology, Atominstitut, TU Wien, Vienna, Austria}
\affiliation{University Service Centre for Transmission Electron Microscopy,TU Wien, Wiedner Hauptstraße 8-10/E057-02, 1040 Wien, Austria}
\author{Peter Schattschneider}
\affiliation{University Service Centre for Transmission Electron Microscopy,TU Wien, Wiedner Hauptstraße 8-10/E057-02, 1040 Wien, Austria}

\date{\today} 

\begin{abstract}


We present an experimental visualization of the Terrell effect, an optical phenomenon predicted in 1959 by Roger Penrose and James Terrell, which reveals that the Lorentz contraction of a moving object is not visible in a snapshot photograph. Using fs-laser pulses and a gated intensified camera that allows gating times as short as 300 ps, we achieve a virtual reduction of the speed of light to less than 2 m/s,  enabling the visualisation of relativistically moving objects in real time. By capturing light reflected from deliberately Lorentz-contracted objects, our setup effectively reconstructs their visual appearance. This didactic visualization not only commemorates the centennial of Anton Lampa's seminal 1924 paper on relativistic length contraction 
but also provides the first experimental  
evidence of the Terrell effect in a laboratory setup. Our results comprise detailed relativistic illustrations, simulations and photographic snapshots of a sphere and a cube, which are animated to velocities close to the speed of light, revealing the apparent rotation effect and the distortion predicted by relativistic theory.

\end{abstract}

\maketitle 
\def\thefootnote{*}\footnotetext{These authors contributed equally to this work}
\newpage
\section{Introduction}
In 1924, Anton Lampa
published a paper on the visual appearance of a relativistically moving rod in Zeitschrift f\"ur Physik\cite{Lampa1924138}. To our knowledge, it is the first discussion of how an observer can effectively measure the relativistic length contraction of a moving rod. Interestingly, the visual appearance of fast-moving 3D objects was not yet an issue at the time.
Only in 1959, Roger Penrose~\cite{Penrose1959137} (and independently James Terrell~\cite{Terrell19591041}) showed that
the Lorentz contraction of a moving object is not visible in a snapshot photo. Under quite general conditions, the moving object appears exactly as the object at rest, but rotated. For example, a sphere always appears as a sphere.   

The effect is due to the fact that for a snapshot photo of a fast-moving object (ideally with exposure time $\rightarrow 0$), light coming from all points of the object must arrive at the same time at the camera. That said, light from more distant points was emitted earlier than light from less distant points. At the earlier time the moving object was at a different position. In the snapshot, the object appears elongated in the direction of movement.
 This compensates the Lorentz contraction - surprisingly, the compensation is exact and the object appears as if at rest, but rotated around an axis orthogonal to the plane of motion and observation. This assumes that the parallel ray projection is valid (
 i. e. when the object is much smaller than the distance to the observer). If this is not the case, distortions are induced, as we discuss later in the text. There are a number of excellent simulations, movies and video games dealing with this phenomenon, e.g.~\cite{Kraus20081}. 

On the occasion of the centennial of Anton Lampa's paper we present a genuine lab experiment that  visualizes the Terrell effect. Based on the photographic technique used by the science-art collective SEEC Photography \cite{deDiosRodríguez}, our experiment provides real-time information on light reflected from the object \cite{duguay1971ultrashortPulse}. 
Our snapshots experimentally demonstrate that the Terrell effect balances the Lorentz contraction and leads to the  appearance of a rotated non-contracted moving object.

\section{Theory}
\begin{figure}[htbp]
	\centering
		\includegraphics[width=\linewidth]{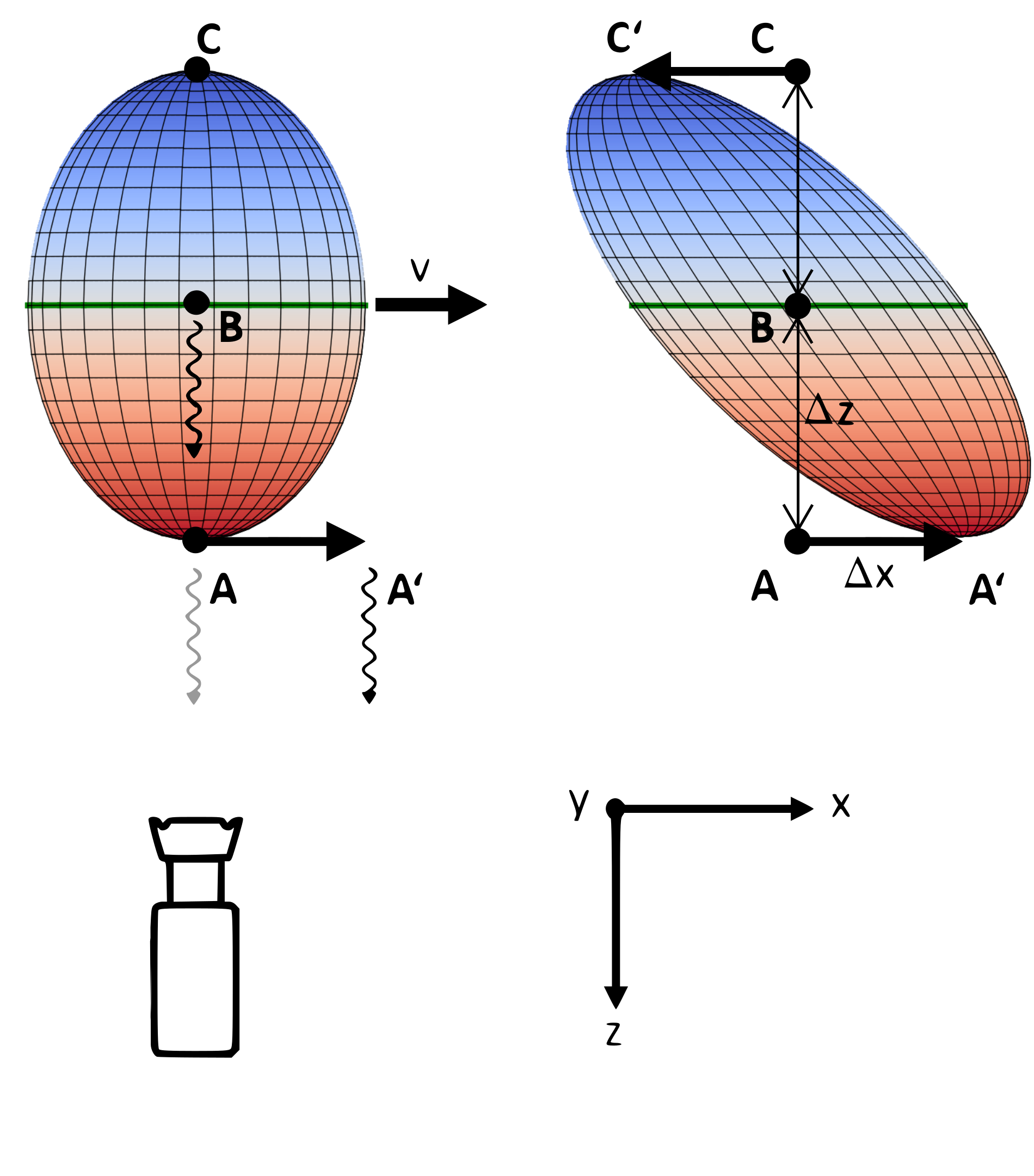}
	\caption{Left: Top view of a  Lorentz contracted sphere moving with speed $v$=0.7 c from left to right. In order that photons from A and B  arrive at the same time at the camera, photon A must be emitted $\Delta t$ later when photon B passes A (grey wiggly line). During this time, point A has moved to position A'.  By symmetry, a photon from point C must be emitted $\Delta t$ {\em earlier} than photon B. In a snapshot, the contracted sphere appears elongated.}
	\label{fig:spheresim}
\end{figure}

\begin{figure}
    \centering
    \includegraphics[width=1\linewidth]{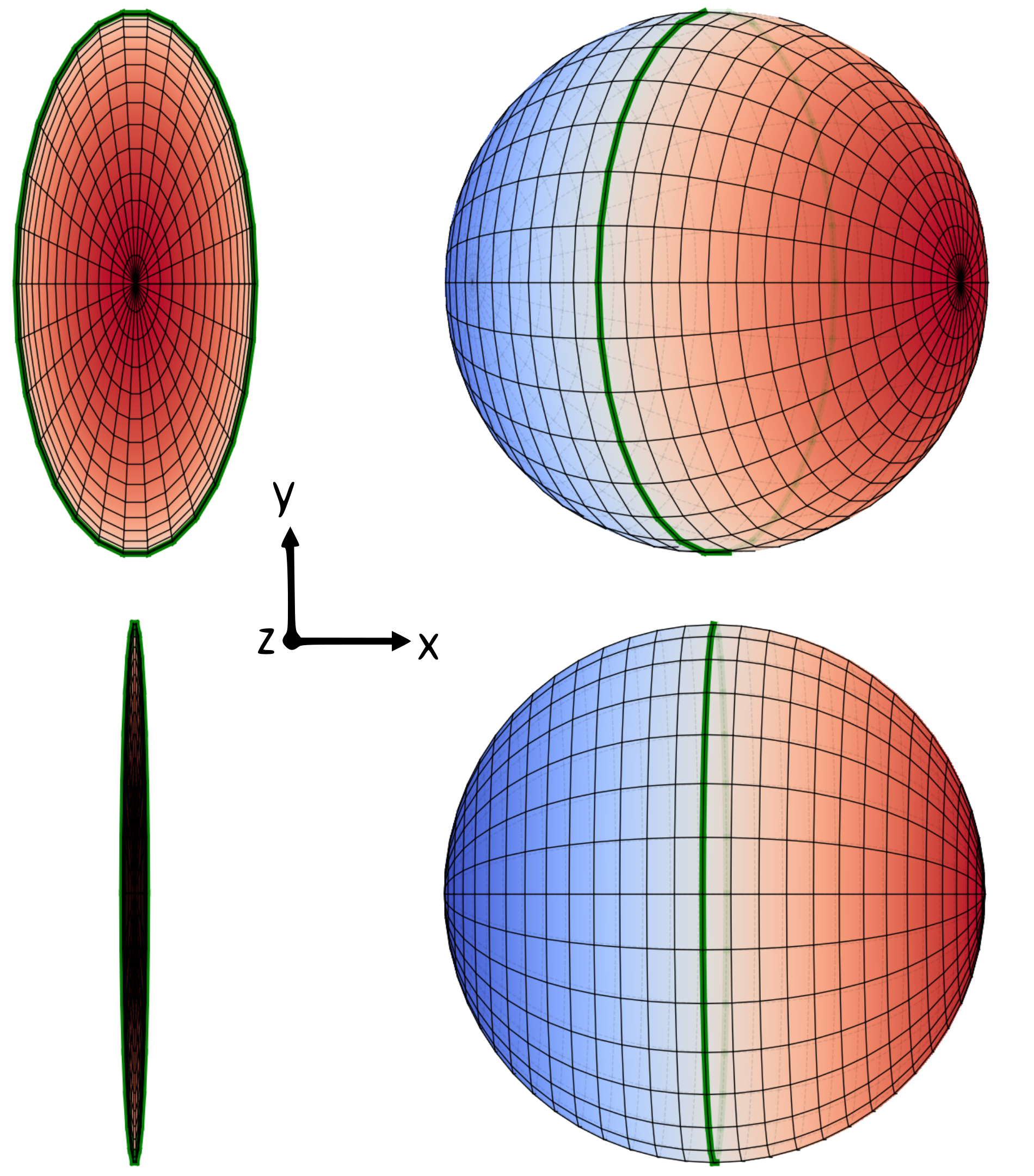}
    \caption{Front view of Fig.~\ref{fig:spheresim} for two different speeds  $v$=0.9, and 0.999~$ c$. Left: Lorentz contracted sphere. The north pole points in direction of the camera. Right: Terrell effect - the sphere appears rotated. The green line indicates the equator.}
    \label{fig:sphere07}
\end{figure}

The principle is illustrated in Fig.~\ref{fig:spheresim}. 
The definition of a snapshot is that photons reflected (or emitted) from any point of the object surface must arrive at the same time at the camera\footnote{In the rest frame of the observer}. Let a photon be emitted from point B at t=0 (wiggly line). In order to arrive at the same time at the camera, a photon from point A (closest to the camera) must be emitted when the photon from B passes point A (grey wiggly line), i.e. at a later time $\Delta t=\Delta z/c$  . During $\Delta t$  point A has moved a distance 
\begin{equation}
\Delta x = v \Delta t=\Delta z v/c
\label{eq:1}
\end{equation}
 to the right to point A'. By symmetry, a photon from point C must be emitted $\Delta t$ {\em earlier} than photon B. Thus, the camera takes a snapshot of an elongated object  as depicted on the right. 
Fig.~\ref{fig:sphere07} shows front views of Fig.~\ref{fig:spheresim} (seen from the camera perspective along the $z$-axis) for different speeds. 
On the left side, the Lorentz contracted sphere is shown, which points its north pole towards the camera/observer. On the right side, we show the points from which the photons creating the snapshot originated. The sphere appears rotated about the  y-axis. One can even look "behind" the equator (green line).

\begin{figure}[h]
    \centering
    \includegraphics[width=0.45\linewidth]{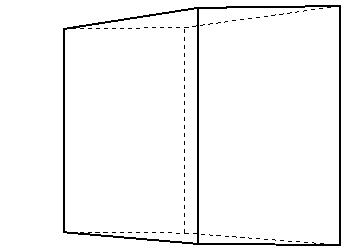}
    \includegraphics[width=0.45\linewidth]{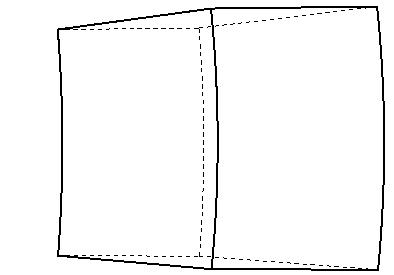}
    \caption{Simulation of the  cube of 1~m side length moving at 0.8~$c$. Left: at a distance of 100 m (almost parallel illumination). Right: at 5 m distance. The  distorted vertical edges are hyperbolae.  }
    \label{fig:cubesim}
\end{figure}

Fig.~\ref{fig:cubesim} is a simulation of a cube moving at 0.8~c from left to right.  As seen on the left, the cube appears rotated; as for the sphere, the Lorentz contraction is not visible. It should be noted that this applies only for parallel ray projections, i.e. for objects with an extension much smaller than their distance to the observer. Otherwise, spherical ray projection causes distortions as shown on the right hand side of Fig.~\ref{fig:cubesim}.

\section{Setup}
To visualize the Terrell effect, we use an ultra-fast photographic technique. We illuminate the object with a pulsed laser and take a snapshot after a certain delay. Light reflected from a specific slice of the object, corresponding to a certain optical path length, will appear bright in this snapshot. The slices are ordered in a temporal sequence.  The delay time between two consecutive exposures defines the distance the real object would move in $x$ direction between two snapshots. The test object is re-positioned accordingly, and the procedure is repeated. The slices are then combined to give the full impression of an object moving at relativistic speeds.

\begin{figure}
    \centering
    \includegraphics[width=\linewidth]{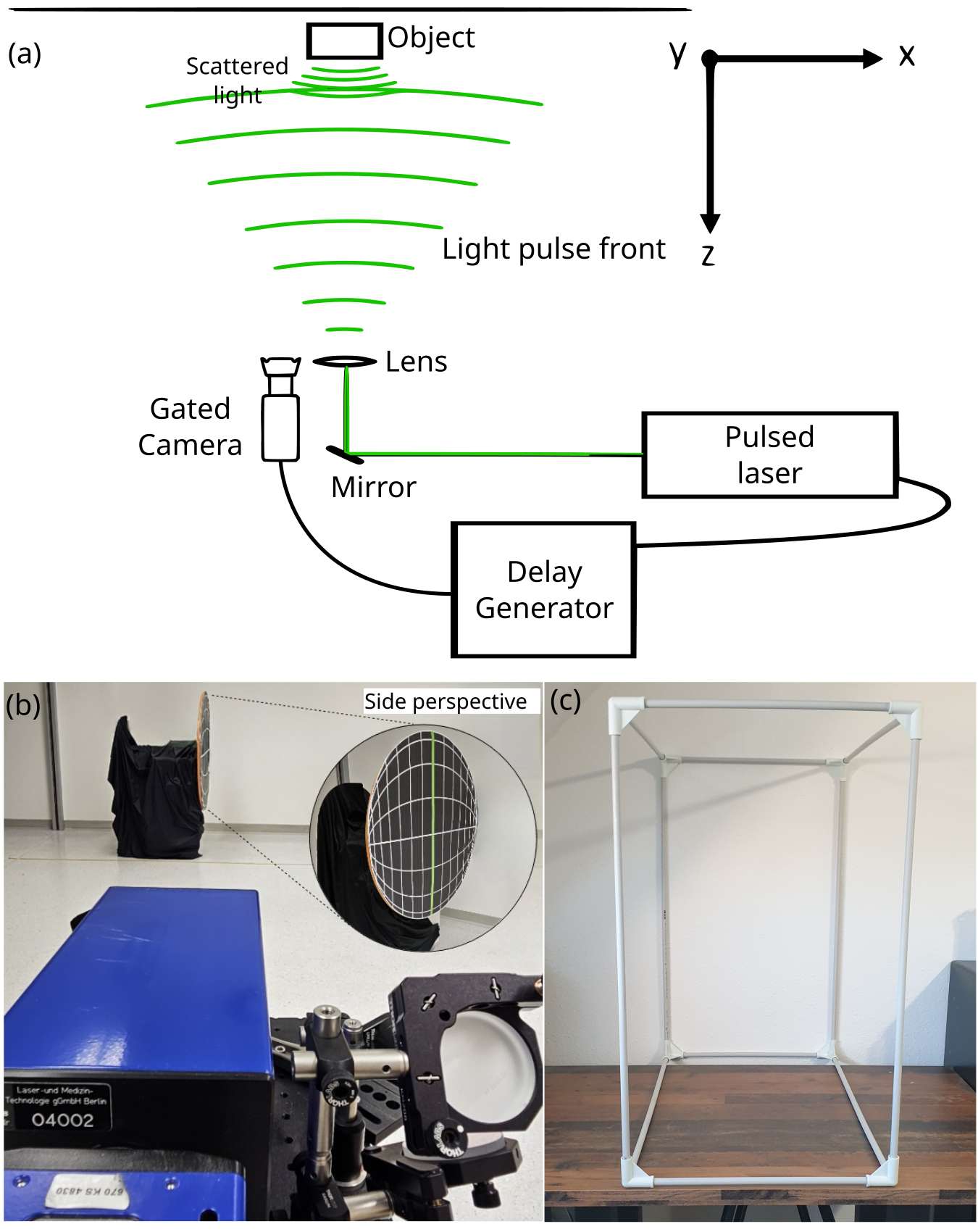}
    \caption{Experimental setup for the snapshot of relativistic motion. (a) The setup consists of a pulsed laser and a gated camera. The laser beam is focused through a lens to illuminate the entire object. The gated camera is triggered to capture the light scattered back from the object with a certain delay in relation to the laser trigger. (b) The slightly tilted model of the Lorentz contracted sphere seen from the camera (blue) for $v=0.999 \, c$ is almost compressed to a 2D object. The north pole points towards the camera.  Next to the camera, the fs-laser pulses are guided via mirrors to a lens in order to expand over the entire field of view. The insert shows the model of the sphere from a side perspective. (c) The Lorentz contracted cube with a side length of $\SI{1}{\meter}$. }
    \label{fig:setup}
\end{figure}

The setup is sketched in Fig.~\ref{fig:setup}a. A $\SI{1035}{nm}$ laser from Coherent labs \footnote{https://www.coherent.com/de/lasers/ultrashort-pulse/monaco} produces 1~ps long laser pulses at a 2~MHz repetition rate. The wavelength of the light is halved to $\SI{517}{nm}$ using a BBO crystal. We create a strongly diverging beam using a lens with a focal length of $\SI{25}{\milli \meter}$. The light scattered off the object is detected by a gated camera (LaVision PicoStar HR12), which is positioned next to the incoming laser beam. The camera and the laser are synchronized using a delay generator.  

The camera is able to detect individual photons by using an image intensifier. The image intensifier is an electronic gating unit enabling gating times as low as $\SI{200}{\pico \second}$, a time span in which the wavefront propagates $\SI{6}{\centi \meter}$.
The High Rate Image Intensifier (Kentech HRI) consists of three main parts: a photo cathode, a microchannel plate (MCP), and a phosphor screen. 
If the gating is off, the photocathode has a positive potential compared to the MCP. The generated photoelectrons are pushed back to the photocathode. If the gating is on, the polarity of the field is changed. The electrons are pushed to the MCP surface, they are multiplied depending on the MCP gain settings (MCP voltage). High electric fields accelerate the electrons towards the phosphor screen, which converts them back to photons that are detected by the CCD camera.

 Two test objects were chosen: a sphere with a diameter of 1 metre at 0.999~c and a cube with an edge length of 1 metre at 0.8~c (see Fig.~\ref{fig:setup}b and Fig.~\ref{fig:setup}c and). Both objects are contracted along the axis of movement by the corresponding Lorentz factor. 
 For this setup a gating time of $\SI{300}{\pico \second}$ was used with an exposure time of $\SI{400}{\milli \second}$. The camera is triggered by the sync output of the laser. For every laser pulse an electronic signal is generated which is accessible through the sync output in the form of rectangular pulses. The delay between laser pulses and camera can be varied through a delay generator (Kentech Instruments HDG800). 
One image series is generated for each position of the object by taking thirty-two pictures with a time delay between single slices of $\SI{400}{\pico \second}$. Light travels $\SI{12}{\centi \meter}$ in that time leading to a spatial distance of $\Delta z=\SI{6}{\centi \meter}$ between two frames. 
After one image series, the object is displaced in the direction of motion. 
For any given speed $v$ the necessary displacement $\Delta x$ between two image series  follows from Eq.~\ref{eq:1} as
$$
\Delta x= \frac{v}{c}{\Delta z}.
$$
For the cube moving at 0.8 c this is  $\SI{48}{\milli \meter}$.

For the sphere seemingly moving at almost the speed of light this corresponds to shifting the object by $\Delta x=\SI{6}{\centi \meter}$ between two series. Note that in order to simulate Lorentz contraction at 0.999~c, the sphere has to be virtually flat.

\begin{figure}[h]
	\centering
		\includegraphics[width=\columnwidth]{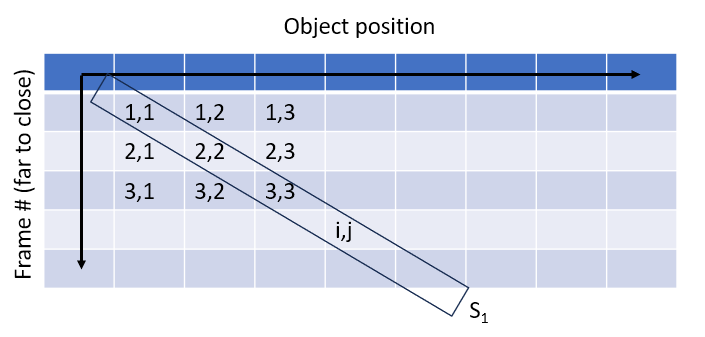}
	\caption{Combining slices of the object to a series of snapshots, i.e. a movie.}
	\label{fig:3}
\end{figure}

With a series of slices $f_{ij}$, where the indices $i$ and $j$ denote timing and position, respectively, a movie can be created by combining slices to snapshots $S_n$ of the moving object as
\begin{equation}
S_n = \{f_{i,i+n},i=1,N\}.
\label{eq:2}
\end{equation}
See the sketch in Fig.~\ref{fig:3}.

After acquiring several slices at different positions, the images were post-processed. To reduce noise from each image, we set low signal pixels to 0. As the strength of genuine photonic signals emanating from scattering objects is at least twice that of background noise, this adjustment has no influence on our results, but leads to a significant reduction of background noise.

To create the image of a fast-moving object, we sum up the slices to a single snapshot $S_1$ according to Eq.~\ref{eq:2} (Fig.~\ref{fig:3}). Each slice is normalized in such a way that the brightest pixel is assigned a value of 1, at the same time accounting for "hot" pixels. This is important to ensure that dark and bright slices are contributing equally. 

Repeating the process for all $S_n$, an animation of a moving object can be created.  At the default rate of 30 frames per second in the slow motion picture, light travels this $\SI{6}{\centi \meter}$ in 1/30~s. The apparent light speed is $c \rightarrow \SI{0.06}{\meter}$/(1/30 s) = $\SI{1.8}{\meter/\second}$.

\section{Results}
\begin{figure}[h]
    \centering
    \includegraphics[width=\linewidth]{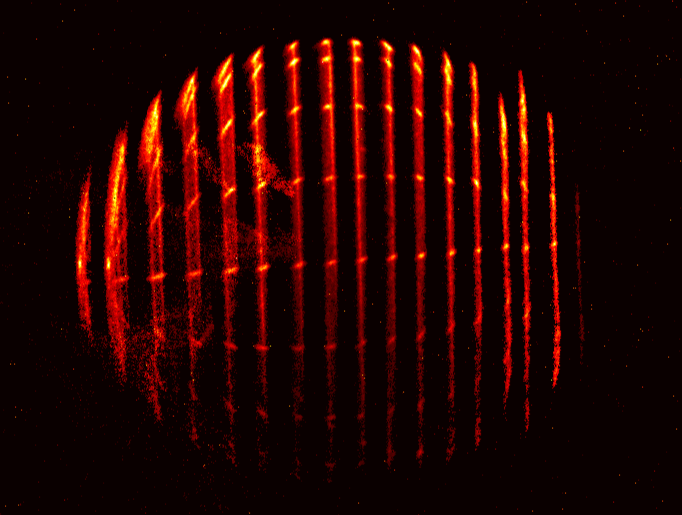}
    \caption{A snapshot of relativistic motion: Experimental data on the Terrell rotation of a deliberately Lorentz contracted sphere at 0.999~c, moving from right to left.}
    \label{fig:sphere}
\end{figure}

 Fig.~\ref{fig:sphere} is a snapshot of the sphere moving from right to left\footnote{For experimental reasons the direction of motion was chosen opposite to the simulation in Figs.~\ref{fig:spheresim} and~\ref{fig:sphere07}} at 0.999~$c$. The north pole that pointed towards the camera as seen in Fig.~\ref{fig:setup}b is located at the leftmost rim. The meridians connecting both poles are visible as brighter dots. Closer inspection 
 shows that the horizontal north-south axis is 11\% longer than the vertical (equatorial) diameter. This effect is an artifact caused by the necessary tilt of the Lorentz-contracted sphere relative to the light source to ensure proper illumination
 . 
The meridians appear discontinuous because they are almost in line of sight seen from the camera. 

\begin{figure}
    \centering
    \includegraphics[width=1\linewidth]{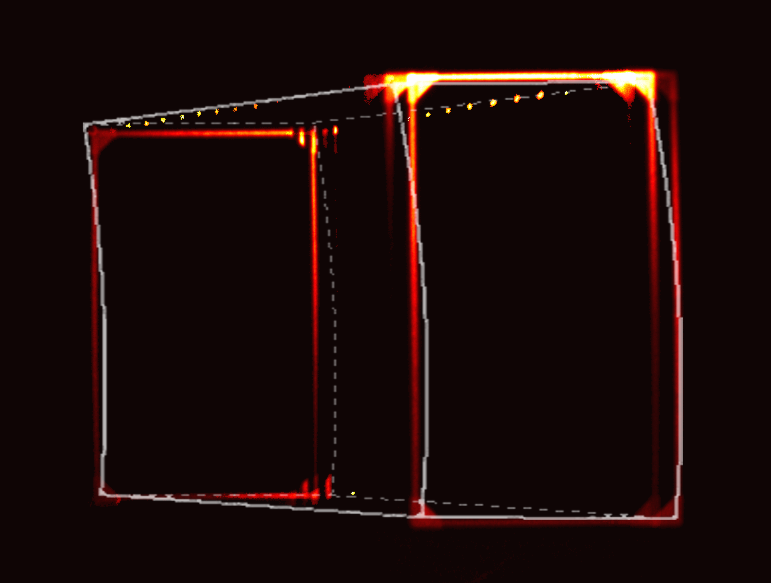}
    \caption{ A snapshot of relativistic motion: Experimental data on the Terrell rotation of a cube. A simulation (white lines) is superimposed on the experimental results to guide the view and verify our theoretical description. The intensity of the double and triple vertical pillars follows the hyperbolic curvature. 
    }
    \label{fig:cubesimclose}
\end{figure}

Fig.~\ref{fig:cubesimclose} is a snapshot of the cube moving from left to right at 0.8~c. As predicted, it appears rotated around the vertical axis.
At first sight, the doubling of the front and back faces and the triplicate left upper front corner are confusing. Comparison  with the simulation (white superimposed frame in Fig.~\ref{fig:cubesimclose}) reveals the cause. The assumption of parallel ray projection is not valid. The inhomogeneous intensity of the edges is caused by the spherical wave emitted by the laser and reflected back to the camera: Not all parts of the bar are visible in the same slice. 
The displacement of the cube  moves parts of the front and back faces into neighbouring slices,  resulting in what appear to be copies. Both effects would cooperate to create the  hyperbolae already shown in Fig.~\ref{fig:cubesim} if the movement were continuous and the gating time infinitely small. 

The bars connecting the front and back sides appear point-like because - similar to the sphere - they are almost in line of sight. Moreover, the  grazing incidence on them back reflects less light into the camera than from the front and back faces.

\section{Conclusion}
This study represents the first experimental visualization of the Terrell effect, initially predicted in 1959 based on Anton Lampa's pioneering work published a century ago. 
Owing to ultrashort gating times down to 0.3 ns \cite{deDiosRodríguez}, a time span in which the pulse moves by less than 10 cm, we were able to demonstrate the invisibility of the Lorentz contraction.
 Our findings confirm the theoretical models, showing that in a snapshot objects moving at relativistic speed appear rotated rather than contracted.
 The detailed visualizations of a sphere and a cube at virtually relativistic speeds align with theoretical expectations, help to understand the so far elusive Terrell effect, and provide intuitive insight into relativistic mechanics.


\section*{Acknowledgments}
We thank Anton Rebhan for valuable comments and Florian Sauer for his support in sample preparation.
PH thanks the Austrian Science Fund (FWF) Y1121, P36041, P35953 and the European Capital of Culture Bad Ischl Salzkammergut 2024. 
TJ acknowledges support from the ERC Micromoupe Grant 758752.







\bibliography{Biblio}
\newpage
\section{Appendix}

\subsection{Simulation of a cube illuminated by a parallel vs. spherical light wave front}

\subsection{Sketch to create the relativistic snapshot}

\subsection{Visual illusion of a rotated sphere}
\begin{figure}[h]
    \centering
    \includegraphics[width=\linewidth]{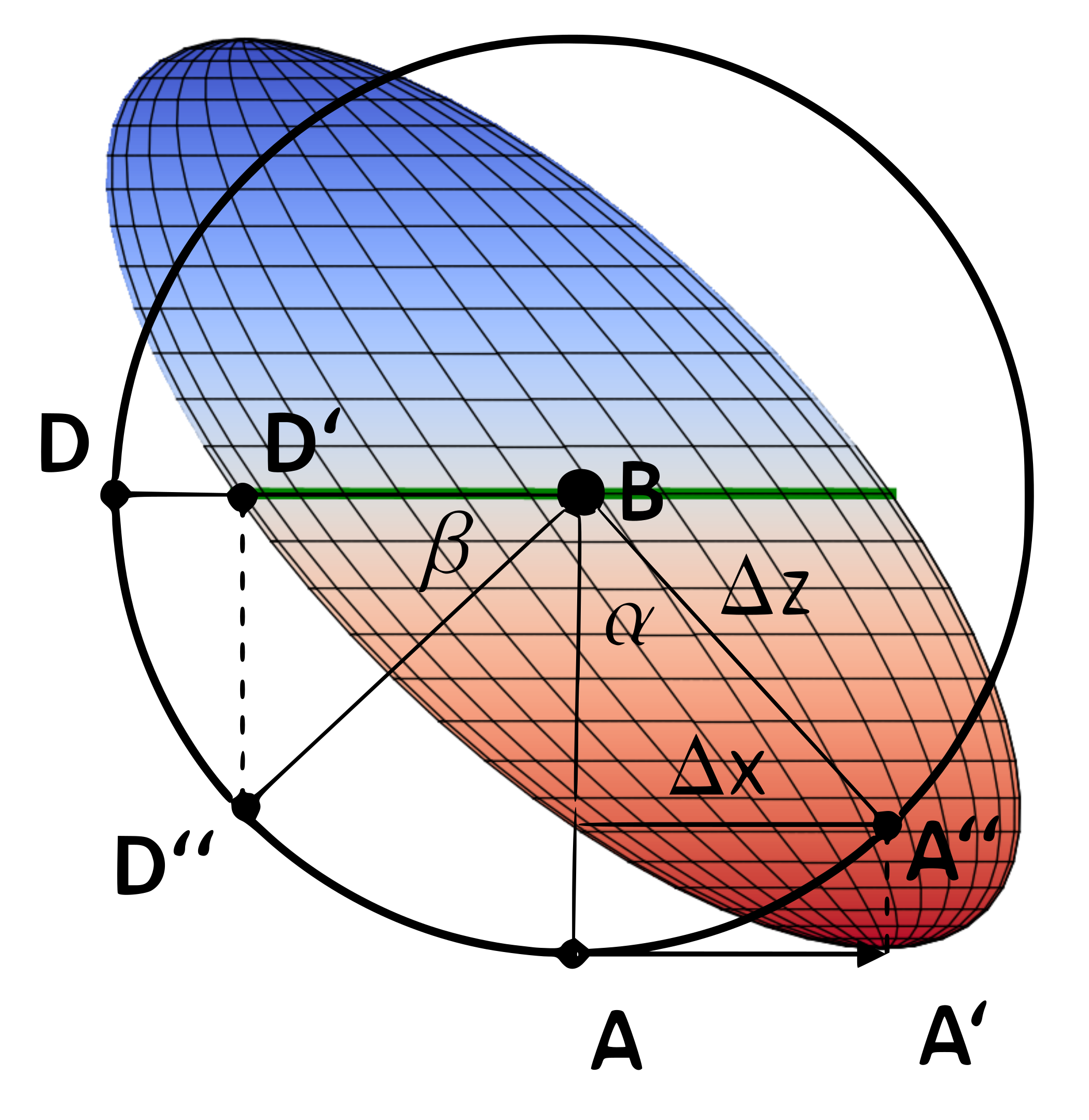}
    \caption{Visual illusion of a rotated sphere under parallel ray projection (dashed lines). Coordinate system as in Fig.\ref{fig:spheresim}}
    \label{fig:rotation}
\end{figure}

Fig.~\ref{fig:rotation}: Point D' on the equator of the  Lorentz contracted sphere of radius $r$ appears as point D'' in parallel ray projection.
$$
\cos\beta=\bar{B D'}/\bar{B D''}=\frac{r/\gamma}{r}=\sqrt{1-(v/c)^2}
$$
whereas the pole A appears as A'' with 
$$
\sin \alpha=\Delta x/\Delta z=v/c.
$$
So
$$
\beta=\alpha.
$$

To the distant observer, the  sheared ellipsoid appears  as a sphere rotated through the angle $\alpha$ about the $y$-axis in the direction of its movement.

\subsection{Animation of the relativistic sphere and cube}
A link to the animations will be provided.

\newpage

\end{document}